\documentclass{ws-ijmpd}

\RequirePackage{graphicx}
\RequirePackage{subfigure}
\RequirePackage{ctable}

\def\bc{\begin{center}}
\def\ec{\end{center}}
\def\be{\begin{eqnarray}}
\def\ee{\end{eqnarray}}

\begin{document}

\markboth{Zhe Chang, Ming-Hua Li, Hai-Nan Lin and Xin Li}
{Generalization of Rindler Potential at Cluster Scales in Randers-Finslerian Spacetime}

%
\catchline{}{}{}{}{}
%

\title{Generalization of Rindler Potential at Cluster Scales in Randers-Finslerian Spacetime: a Possible Explanation of the Bullet Cluster 1E0657-558 ?}

\author{Zhe Chang}

\address{Institute of High Energy Physics, Chinese Academy of Sciences, 100049 Beijing, China, and\\
Theoretical Physics Center for Science Facilities, Chinese Academy of Sciences, 100049 Beijing, China\\
changz@ihep.ac.cn}

\author{Ming-Hua Li}

\address{Institute of High Energy Physics, Chinese Academy of Sciences, 100049 Beijing, China\\
limh@ihep.ac.cn}

\author{Hai-Nan Lin}

\address{Institute of High Energy Physics, Chinese Academy of Sciences, 100049 Beijing, China\\
limh@ihep.ac.cn}

\author{Xin Li}

\address{Institute of High Energy Physics, Chinese Academy of Sciences, 100049 Beijing, China, and\\
Theoretical Physics Center for Science Facilities, Chinese Academy of Sciences, 100049 Beijing, China\\
lixin@ihep.ac.cn}

\maketitle

\begin{history}
\received{Day Month Year}
\revised{Day Month Year}
\comby{Managing Editor}
\end{history}

\begin{abstract}
The data of the Bullet Cluster 1E0657-558 released on November 15, 2006 reveal that the strong and weak gravitational lensing convergence $\kappa$-map has an $8\sigma$ offset from the $\Sigma$-map. The observed $\Sigma$-map is a direct measurement of the surface mass density of the Intracluster medium(ICM) gas. It accounts for $83\%$ of the averaged mass-fraction of the system. This suggests a modified gravity theory at large distances different from Newton's inverse-square gravitational law. In this paper, as a cluster scale generalization of Grumiller's modified gravity model (D. Grumiller, Phys. Rev. Lett. 105, 211303 (2010)), we present a gravity model with a generalized linear Rindler potential in Randers-Finslerian spacetime without invoking any dark matter. The galactic limit of the model is qualitatively consistent with the MOND and Grumiller's. It yields approximately the flatness of the rotational velocity profile at the radial distance of several kpcs and gives the velocity scales for spiral galaxies at which the curves become flattened. Plots of convergence $\kappa$ for a galaxy cluster show that the peak of the gravitational potential has chances to lie on the outskirts of the baryonic mass center. Assuming an isotropic and isothermal ICM gas profile with temperature $T=14.8$ keV (which is the center value given by observations), we obtain a good match between the dynamical mass $M_\textmd{T}$ of the main cluster given by collisionless Boltzmann equation and that given by the King $\beta$-model. We also consider a Randers$+$dark matter scenario \textbf{and a $\Lambda$-CDM model} with the NFW dark matter distribution profile. We find that a mass ratio $\eta$ between dark matter and baryonic matter about 6 fails to reproduce the observed convergence $\kappa$-map for the isothermal temperature $T$ taking the observational center value.
\end{abstract}

\keywords{Modified Gravity; Rindler potential; Finsler geometry; Randers spacetime; Bullet Cluster.}

\section{Introduction}

It has long been known that the gravitational potentials of some galaxy clusters are too deep to be generated by the observed baryonic matter according to Newton's inverse-square law of gravitation \cite{Zwicky1933}. This violation of Newton's law is further confirmed by a great variety of observations. To name a few: the Oort discrepancy in the disk of the Milky Way \cite{Bahcall1992}, the velocity dispersions of dwarf Spheroidal galaxies \cite{Vogt1995}, and the flat rotation curves of spiral galaxies \cite{Rubin1980}.
The most widely adopted way to solve these mysteries is to assume that all our galaxies and clusters are surrounded by massive non-luminous dark matter \cite{Oort1932}. Despite its phenomenological success in explaining the flat rotation curves of spiral galaxies, the hypothesis has its own deficiencies. No theory predicts these matters, and they behave in such ad hoc way like existing as a halo without undergoing gravitational collapse. There are a lot of possible candidates for dark matter (such as axions, neutrinos {\it et al.}), but none of them are sufficiently satisfactory. Up to now, all of them are either undetected or excluded by experiments and observations.

Because of all these troubles, some models have been built as alternatives of the dark matter hypothesis. Their main ideas are to suggest that Newton's dynamics is invalid in the galactic scale.
A famous example is the MOND \cite{Milgrom1983}.
It supposes that in the galactic scale, the Newton's dynamics appears as
\begin{equation}
\begin{array}{l}
\label{MOND}
m\mu\left(\displaystyle\frac{a}{a_0}\right)\mathbf{a}=\mathbf{F},\\[0.4cm]
\displaystyle\lim_{x\gg1}\mu(x)=1,~~~\lim_{x\ll1}\mu(x)= x,
\end{array}
\end{equation}
where $a_0$ is a constant and the value of which is of order $10^{-8}$ cm/s$^2$. Dwarf and low surface brightness galaxies provide a good test for the MOND \cite{Sanders2002}. With a simple formula and the one-and-only-one constant parameter $a_0$, the MOND yields the observed luminosity-rotation velocity relation, the Tully-Fisher relation \cite{Tully1977}. By introducing several scalar, vector and tensor fields, Bekenstein developed a relativistic version of the MOND \cite{Bekenstein2004}. The covariant MOND satisfies all four classical tests on Einstein's general relativity in Solar system.

Although the MOND successfully reduces the discrepancy between the visible and the Newtonian dynamical mass (which is also quantified in
terms of mass-to-light ratio) to a factor of $2/3$, there still remains a missing mass problem, particularly in the cores of clusters of galaxies \cite{Sanders2003}. The data release of the Bullet Cluster 1E0657-558 in November of 2006 posed a serious challenge for modified gravity theories such as the MOND.

The Bullet Cluster 1E0657-558 was first spotted by the Chandra X-ray Observatory in 2002 \cite{Markevitch2002}. Located at a redshift $z = 0.296$ (Gpc scale), it has exceptionally high X-ray luminosity and is one of the largest and hottest luminous galaxy clusters in the sky. A high-resolution map of the ICM gas, i.e. the surface mass density $\Sigma(x,y)$, was reconstructed by Clowe {\it et al.} \cite{Clowe2006c,Clowe2006b} in 2006. It exhibits a supersonic shock front in the plane of the merger, which is just aligned with our sky. The high-resolution and absolutely calibrated convergence $\kappa$-map of the sky region that surrounds the ``bullet'' was also reconstructed by Brada\v{c} and Clowe {\it et al.} in their gravitational lensing surveys \cite{Bradac2006,Clowe2004,Clowe2006a}.
The $\kappa$-map is evidently offset from the $\Sigma$-map. The peak of the $\kappa$-map lies on the region of galaxies instead of tracing the ICM gas of the main cluster, which makes up about $83\%$ of the total baryonic mass of the merging system.

Clowe {\it et al.} \cite{Clowe2006c,Clowe2004,Clowe2006a} took it as a direct empirical evidence of the existence of dark matter, while whether the MOND could fit the X-ray temperature profiles without dark matter component is still in issue \cite{Sanders2003,Aguirre2001,Angus2006a,Angus2006b,Takahashi2007}. Using their modified gravity (MOG), Brownstein and Moffat partly explained the steepened peaks of the $\kappa$-map, while attributing the rest differences to the MOG's effect of the galaxies \cite{Brownstein2007}.

On the other hand, Grumiller \cite{Grumiller2010} presented an effective model for gravity of a central object at large scales recently. To leading order in the large radius expansion, the action of his model leads to an additional ``Rindler term'' in the gravitational potential. This extra term gives rise to a constant acceleration towards or away from the source.
The scale where the velocity profile flattens is $v\sim 300$ km/s, in reasonable agreement with the observational data.

In this paper, inspired by these prominent work, we try to construct a modified gravity model at large distances with a generalized Rindler potential without invoking any dark matter. This is carried out in a Randers-Finslerian spacetime in Zermelo's navigation scenario \cite{Bao2004,Gibbons2009,Zermelo1931}. Finslerian geometry is a generalization of Riemannian geometry without quadratic restrictions on the line element \cite{Chern1995}. It is intriguing to investigate the possible physical implication in such a general geometrical background. In fact, precedent work have yielded some interesting results \cite{Stavrinos2008,Chang2008,Chang2009a,Chang2009b,Chang2010a,Chang2010b,Asanov1985,Bogoslovsky1993,Ikeda1987,Takano1974,Tavakol1985,Vacaru2008,Stavrinos2005,Kouretsis2010}. 
The work in this paper is a cluster-scale generalization of Grumiller's model and it is ensured that in the galactic limit, it agrees with both the Grumiller's model and the observational data. An approximately flattened velocity profile predicted by our model makes it qualitatively consistent with the MOND at the distance scale of several kpcs. The Newtonian limit and the gravitational deflection of light are particularly investigated and the deflection angle is given explicitly.

We use the isothermal King $\beta$-model to describe the observed $\Sigma$-map of a galaxy main cluster. The convergence $\kappa$ is obtained. It is found that the gravitational potential peak does not always lie on the center of the baryonic material center. Chances are that it will has a bigger value in the outskirts rather than the center. This is one of the distinguishing features of the reconstructed $\kappa$-map of the Bullet Cluster system. Besides, the gravity provided by the baryonic material is somehow ``enlarged''.
It is reasonable to suggest that these results may ameliorate the conundrum between the gravity theory and the observations of the Bullet Cluster 1E0657-558.

The rest of the paper is organized as follows. Section 2 is divided into four parts: in Section 1, we introduce the basic concepts of Finsler geometry; in Section 2.1, we use the the second Bianchi identities to get the gravitational field equation in Berwald-Finslerian space; in Section 2.3, we consider a Randers-type spacetime in a navigation scenario with a vector field in the radial direction; in Section 2.4, we integrate the geodesic equation to get the deflection angle in Randers-Finslerian spacetime with a generalized Rindler potential at cluster scales. Section 3 is divided into five parts: in Section 3.1, we give the Poisson's equation by which the effective lens potential obeys; in Section 3.2, by making use of the effective lens potential, we obtain the convergence $\bar{\kappa}$ of the Bullet Cluster 1E0657-558. The cross section of the calculated $\bar{\kappa}$-map is presented; in Section 3.3, the isothermal temperature of the main cluster is calculated; in Section 3.4, we consider a Randers$+$dark matter model for comparison; in Section 3.5, we investigate the performance of our model at galactic scales. Conclusions and discussions are presented in Section 4. Appendix is in the last section.

\section{Finslerian Geometry}

\subsection{Basic Concepts}
Finslerian geometry is a natural generalization of Riemannian geometry without quadratic restrictions on the metric \cite{Chern1995}. It is based on a real function $F$ called Finsler structure (or Finslerian norm in some literature) with the property $F(x, \lambda y)=\lambda F(x, y)$ for all $\lambda>0$, where $y^{\mu} \equiv dx^{\mu}/{d\tau}$ ($\mu = 0, 1, 2,~...~, n$). In physics, $x^{\mu}$ stands for position and $y^{\mu}$ stands for velocity. The metric of Finslerian space is given by \cite{Bao2000}
\be
g_{\mu\nu}\equiv\frac{\partial}{\partial y^\mu}\frac{\partial}{\partial y^\nu}\left(\frac{1}{2}F^2\right)\,.
\label{g}
\ee

Finslerian geometry has its genesis in the integral of the form
\be
\label{integral length}
\int^r_s F(x^1,\cdots,x^n;y^1,\cdots,y^n)d\tau~\, .
\ee
It represents the arc length of a curve in a Finslerian manifold.
The first variation of (\ref{integral length}) gives the geodesic equation in a Finslerian space \cite{Bao2000}
\be
\label{geodesic0}
\frac{d^2x^\mu}{d\tau^2}+G^\mu=0\ ,
\ee
where
\be
\label{geodesic spray}
G^\mu\equiv\frac{1}{2}g^{\mu\nu}\left(\frac{\partial^2 F^2}{\partial x^\lambda \partial y^\nu}y^\lambda-\frac{\partial F^2}{\partial x^\nu}\right)
\ee
is called the geodesic spray coefficient. Obviously, if $F$ is Riemannian metric, then
\begin{equation}
G^\mu=\tilde{\gamma}^\mu_{~\nu\lambda}y^\nu y^\lambda,
\end{equation}
where $\tilde{\gamma}^\mu_{~\nu\lambda}$ is the Riemannian Christoffel symbol.

In a Finslerian manifold, there exists a unique linear connection~-~the
Chern connection \cite{Chern1948}. It is torsion freeness and almost
metric-compatibility,
 \begin{equation}
 \label{Chern connection}
\Gamma^{\alpha}_{~\mu\nu}=\gamma^{\alpha}_{~\mu\nu}-g^{\alpha\lambda}\left(A_{\lambda\mu\beta}\frac{N^\beta_{~\nu}}{F}-A_{\mu\nu\beta}\frac{N^\beta_{~\lambda}}{F}+A_{\nu\lambda\beta}\frac{N^\beta_{~\mu}}{F}\right),
 \end{equation}
 where $N^\mu_{~\nu}$
is defined as
$N^\mu_{~\nu} \equiv\gamma^\mu_{~\nu\alpha}y^\alpha-A^\mu_{~\nu\lambda}\gamma^\lambda_{~\alpha\beta}y^\alpha y^\beta$
 and $A_{\lambda\mu\nu}\equiv\frac{F}{4}\frac{\partial}{\partial y^\lambda}\frac{\partial}{\partial y^\mu}\frac{\partial}{\partial y^\nu}(F^2)$ is the
 Cartan tensor (regarded as a measurement of deviation from the Riemannian
 Manifold).
In terms of Chern connection, the curvature of Finsler space is given as
\begin{equation}
\label{Finsler curvature}
R^{~\lambda}_{\kappa~\mu\nu}=\frac{\delta
\Gamma^\lambda_{~\kappa\nu}}{\delta x^\mu}-\frac{\delta
\Gamma^\lambda_{~\kappa\mu}}{\delta
x^\nu}+\Gamma^\lambda_{~\alpha\mu}\Gamma^\alpha_{~\kappa\nu}-\Gamma^\lambda_{~\alpha\nu}\Gamma^\alpha_{~\kappa\mu},
\end{equation}
where $\frac{\delta}{\delta x^\mu}=\frac{\partial}{\partial x^\mu}-N^\nu_{~\mu}\frac{\partial}{\partial y^\nu}$.

\subsection{Field Equations}
Constructing a physical Finslerian theory of gravity in an arbitrary Finslerian spacetime is a difficult task.
However, it has been pointed out that constructing a Finslerian theory of gravity in a Finlserian spacetime of Berwald type is viable \cite{Tavakol1985}.
A Finslerian spacetime is said to be of Berwald type if the Chern connection (\ref{Chern connection}) have no $y$ dependence\cite{Bao2000}.
In the light of the research of Tavakol {\it et al}. \cite{Tavakol1985}, the gravitational field equation in Berwald-Finslerian space has been studied in \cite{Chang2008,Chang2010b}. In Berwald-Finslerian space, the Ricci tensor reduces to
\begin{equation}
Ric_{\mu\nu}=\frac{1}{2}(R^{~\alpha}_{\mu~\alpha\nu}+R^{~\alpha}_{\nu~\alpha\mu})\, .
\end{equation}
It is manifestly symmetric and covariant. Apparently it will reduce to the Riemann-Ricci tensor if the metric tensor $g_{\mu\nu}$ does not depend on $y$. We starts from the
second Bianchi identities in Berwald-Finslerian space \cite{Bao2000}
\begin{equation}
\label{Bianchi on Riemann} R^{~\alpha}_{\mu~\lambda\nu|\beta}+R^{~\alpha}_{\mu~\nu
\beta|\lambda}+R^{~\alpha}_{\mu~\beta\lambda|\nu}=0\ ,
\end{equation} where the
``$|$'' means the covariant derivative.
The metric-compatibility $g_{\mu\nu|\alpha}=0$ and $g^{\mu\nu}_{~~~|\alpha}=0$ and contraction of (\ref{Bianchi on Riemann}) with $g^{\mu\beta}$
gives that
\begin{equation} R^{\mu
\alpha}_{~~~\lambda\nu|\mu}+R^{\mu\alpha}_{~~~\nu\mu|\lambda}+R^{\mu\alpha}_{~~~\mu\lambda|\nu}=0\, .
\end{equation}
Lowering the index $\alpha$ and contracting with
$g^{\alpha\lambda}$, we obtain
\begin{equation}
\left[Ric_{\mu\nu}-\frac{1}{2}g_{\mu\nu}S\right]_{|\mu}+\left\{\frac{1}{2}B^{~\alpha}_{\alpha~\mu\nu}+B^{~\alpha}_{\mu~\nu\alpha}\right\}_{|\mu}=0
,\end{equation} where
\be
B_{\mu\nu\alpha\beta}&=&-A_{\mu\nu\lambda}R^{~\lambda}_{\theta~\alpha\beta}y^\theta/F\ ,\\
R&=&\frac{y^\mu}{F}R^{~\kappa}_{\mu~\kappa\nu}\frac{y^\nu}{F}\ ,\\
S&=&g^{\mu\nu}Ric_{\mu\nu}\, .
\ee
 Thus, we get the counterpart of the Einstein's field equation in
 Berwald~-~Finslerian space
 \begin{equation}\label{berwald field eq}
\left[Ric_{\mu\nu}-\frac{1}{2}g_{\mu\nu}S\right]+\left\{\frac{1}{2}B^{~\alpha}_{\alpha~\mu\nu}+B^{~\alpha}_{\mu~\nu\alpha}\right\}=8\pi
G T_{\mu\nu}\, .
\end{equation}
In Eq. (\ref{berwald field eq}), the term in ``[~]" is symmetrical tensor, and the term in ``\{\}" is asymmetrical tensor.
By making use of Eq. (\ref{berwald field eq}), the vacuum field equation in Finslerian spacetime of Berwald type implies
\begin{equation}
Ric_{\mu\nu}=\frac{1}{2}(R^{~\alpha}_{\mu~\alpha\nu}+R^{~\alpha}_{\nu~\alpha\mu})=0\, .
\end{equation}

\subsection{Randers type space with a ``Wind''}
Randers space is a special kind of Finslerian geometry with the Finsler structure $F$ defined on the slit tangent bundle $TM\backslash0$ of a manifold $M$ as \cite{Bao2000,Randers1941},
\be
F(x,y) = \alpha(x,y)+\beta(x,y)\ ,
\label{Randers}
\ee
where
\begin{eqnarray}
\alpha(x,y) &\equiv& \sqrt{\tilde{a}_{\mu \nu}(x)y^{\mu} y^{\nu} }\ ,\\
\beta(x,y) &\equiv& \tilde{b}_{\mu}(x)y^{\mu}\ .
\end{eqnarray}
Here, $\tilde{a}_{\mu\nu}$ is a Riemannian metric and $\tilde{b}_\mu$ is an 1-form. Here and after, if not specified, lower case Greek indices (i.e. $\mu, \nu, \alpha, ...$) run from $0$ to $3$ and the Latin ones (i.e. $i, j, k, ...$) run from $1$ to $3$.
Positivity of $F$ holds if and only if \cite{Bao2000}
\be
|\tilde{b}|\equiv\sqrt{\tilde{b}_\mu \tilde{b}^\mu}~~<1\ ,
\label{b condition}
\ee
where
\be
\tilde{b}^\mu \equiv \tilde{a}^{\mu\nu} \tilde{b}_\nu\, .
\ee

Stavrinos \textit{et al.} \cite{Stavrinos2008} constructed a generalized FRW model based on a Lagrangian identified to be the Randers-type metric function. New Friedman equations and a physical generalization of the Hubble and other cosmological parameters were obtained. Zermelo \cite{Zermelo1931} aimed to find minimum-time trajectories in a Riemannian manifold $(M, h)$ under the influence of a ``wind'' represented by a vector field $W$ . Shen \cite{Shen2003} proved that the minimum time trajectories are exactly the geodesics of Randers space, if the wind is time independent.

In this paper, we consider a Randers-Finslerian structure $F(x,y)$ under the influence of a ``wind'' in the radial direction $W\equiv W_{\mu}dx^\mu = W_r dr$, to wit
\be
\tilde{a}_{\mu\nu} = \frac{\lambda h_{\mu \nu}+ W_\mu W_\nu}{\lambda^2},~~\tilde{b}_{\mu} = -\frac{W_\mu}{\lambda},~~\lambda = 1-h_{\mu \nu}W^\mu W^\nu\ ,
\label{Randers2}
\ee
where $W^\mu = h^{\mu \nu}W_{\nu}$ and $\tilde{a}^{\mu\nu} = \lambda (h^{\mu \nu}- W^\mu W^\nu)$.
Here $h_{\mu \nu}$ is the Schwarzschild metric
\be
h_{ij}dx^i dx^j =\left(1- \frac{2GM}{r}\right)^{-1}dr^2 +r^2 d\theta ^2 +r^2 sin^2 \theta d\varphi ^2\, .
\label{Schwarzschild}
\ee

From (\ref{Randers2}), we have
\be
\tilde{b}_r=-\frac{1}{\lambda} \sqrt{ \frac{1-\lambda}{\left(1- \frac{2GM}{r}\right)^{-1}}}\ ,
\ee
where $\lambda$ is a function of $r$, i.e. $\lambda=\lambda(r)$.
Zermelo \cite{Zermelo1931} said little about the $\lambda$ except for the condition that the size of the component $\tilde{b}_{r}$ must be suitably controlled, i.e. $|\tilde{b}_r|<1$, for $F$ to be positive on $TM\backslash0$. But for a physical model, the specific form of $\lambda(r)$ is determined not only by the local symmetry of the spacetime but also constrained by the experiments and observations.

The explicit form of $F(x,y)$ reads
\begin{eqnarray}
F d\tau &=& \sqrt{ \lambda ^{-1} \left( \left( 1- \frac{2GM}{r} \right) ^{-1} dr^2 + r^2 d\theta ^2 + r^2 sin^2 \theta d\varphi ^2 \right) + \lambda^{-2} W_{r}^2 dr^2} - \lambda^{-1} W_{r} dr \nonumber \\
&=& \sqrt{\lambda ^{-2}\left(1- \frac{2GM}{r}\right)^{-1}dr^2 + \lambda ^{-1}\left(r^2 d\theta ^2 +r^2 sin^2 \theta d\varphi ^2\right)} - \lambda^{-1}W_{r} dr\ ,
\label{nonrelativisticF}
\end{eqnarray}
where the second equation exploits the expression of $\lambda$ in (\ref{Randers2}) assuming $|\frac{GM}{r}|\ll 1$.
The relativistic form of (\ref{nonrelativisticF}) is given as \footnote{In Chapter 8 of \cite{Weinberg1972}, the standard form of the proper time interval of a static isotropic or approximately static isotropic gravitational field is given as
\be
d\tau ^2= B(r)dt^2 - A(r)dr^2 -r^2 \left(d\theta ^2 + sin^2 \theta d\varphi ^2\right)\, . \nonumber
\ee
The field equations for empty space $R_{\mu \nu}=0$ requires that $A(r)B(r)= \textmd{constant}$. And the metric tensor must approach the Minkowski tensor in spherical coordinates, that is, for $r\rightarrow \infty$, $A(r)=B(r)=1$. Thus we have
\be
A(r)B(r)=1\, . \nonumber
\ee
For the Randers-Finslerian metric (\ref{Randers}) and (\ref{nonrelativisticF}), that is
\be
\tilde{a}_{00}= -\lambda ^{2}\left(1- \frac{2GM}{r}\right)\, . \nonumber
\ee
}
\be
F d\tau =  \sqrt{-\lambda ^{2}\left(1- \frac{2GM}{r}\right)dt^2 + \lambda ^{-2}\left(1- \frac{2GM}{r}\right)^{-1}dr^2 + \lambda ^{-1}\left(r^2 d\theta ^2 +r^2 sin^2 \theta d\varphi ^2\right)} - \lambda^{-1}W_{r} dr\ .
\label{relativisticF}
\ee
Discussions in the next subsection are based on the geodesic equation which stems from a Lagrangian identified to be the Randers-type metric function (\ref{relativisticF}) in four-dimensional spacetime.

\subsection{Equations of Montion and Deflection Angle}
In a Randers space, the geodesic equation (\ref{geodesic0}) takes the form of
\footnote{We just consider the case that the $\beta$ in (\ref{Randers}) is a closed 1-form, i.e. $d\beta =0$.}
\be
\frac{d^2x^\mu}{d\tau^2}+ \left(\tilde{\gamma}^{\mu}_{~\nu\alpha}+ \ell^{\mu} \tilde{b}_{\nu|\alpha} \right) y^{\nu} y^{\alpha}=0\ ,
\label{geodesics}
\ee
where
\be
\ell^{\mu}\equiv\frac{y^\mu}{F},~~~~\tilde{b}_{\nu|\alpha}\equiv\frac{\partial\tilde{b}_\nu}{\partial
x^\alpha}-\tilde{\gamma}^\mu_{~\nu\alpha}\tilde{b}_\mu \ ,
\ee
and $\tilde{\gamma}^{\mu}_{~\nu\alpha}$ is the Christoffel symbols of the Riemannian metric $\tilde{a}_{\mu \nu}$.
Given the Finslarian structure in (\ref{relativisticF}), the non-vanishing components of the geodesic equations (\ref{geodesics}) (i.e. the equation of motion) give rise to the relation between the radial distance $r$ and the angle $\varphi$ of the orbits of free particles, to wit \cite{Li2011}
\be
\left(\frac{1}{r^2}\frac{dr}{d\varphi}\right)^2=\left(\frac{E}{J\lambda(r)}\right)^2-\frac{\lambda(r)}{r^2}\left(1-\frac{2GM}{r}\right)\ ,
\label{bend light}
\ee
where $E$ and $J$ are the integral constants of motion. Introducing a new quantity
\be
u \equiv \frac{GM}{r}\ ,
\label{u}
\ee
Eq. (\ref{bend light}) can be rewritten in terms of $u$ as
\be
\label{eq u}
\left(\frac{du}{d\varphi}\right)^2=\left(\frac{EGM}{J\lambda(r)}\right)^2-\lambda u^2(1-2u)\, .
\ee
It should be noticed that the only difference between Eq. (\ref{eq u}) and its Riemmanian counterpart is the $\lambda(r)$. For $\lambda\rightarrow 1$, Eq. (\ref{eq u}) returns to that in the general relativity.

To describe a real physical system, one has to give a specific form of $\lambda(r)$.
In this paper we consider
\be
\label{lambda}
\lambda(r) = 1- \frac{GM}{r_s}\left(1+\frac{r}{r_e}\right)e^{-\frac{r}{r_e}}\, .
\ee
$r_s$ and $r_e$ parameterize the physical scales of the system.
As we stated before, one of the restrictions for $\lambda(r)$ is to ensure that $|\tilde{b}_r|<1$. It is shown in Section 4 that (\ref{lambda}) satisfies this condition.
Given (\ref{lambda}), one can solve \footnote{See the Appendix for details.} the equation of motion (\ref{eq u}), which is derived from (\ref{relativisticF}). The result is
\begin{equation}\label{modified potential}
\phi_{M}=-\frac{GM}{r}-\frac{GM}{r_s}\left(1+\frac{r}{r_e}\right)e^{-\frac{r}{r_e}}\, .
\end{equation}
The first term in (\ref{modified potential}) is the usual Newtonian potential and the last linear term with an exponential cutoff is novel.

The particular function form (\ref{lambda}) of the parameter $\lambda$ is inspired by Grumiller's work \cite{Grumiller2010}. The effective potential in his paper was given as
\begin{equation}\label{modified potential1}
\phi_M=-\frac{GM}{r}+Dr\ ,
\end{equation}
where $D$ is constant and the linear term $Dr$ is called the Rindler acceleration term. A more general form of (\ref{modified potential1}) can be written as
\begin{equation}\label{modified potential2}
\phi_M=-\frac{GM}{r}+\tilde{f}(r)\ ,
\end{equation}
where $\tilde{f}(r)$ is a function of the distance scale $r$. For Grumiller's model, $\tilde{f}(r)=Dr$. And for the specific form of $\lambda(r)$ in (\ref{modified potential}), $\tilde{f}(r)$ takes a form as
\begin{equation}\label{ff}
\tilde{f}(r)=-\frac{GM}{r_s}\left(1+\frac{r}{r_e}\right)e^{-\frac{r}{r_e}}\, .
\end{equation}
It is a rescaled linear potential of $r$ like the Grumiller's, but an exponential cutoff is imposed to avoid possible divergence at large distances. Grumiller's potential does not confront with such a difficulty because he only discussed the galactic physics. While we try to extrapolate the potential (\ref{modified potential1}) to the cluster scale, we do need to consider this problem.
The effective acceleration $a_{M}$ has two terms, also
\be
a_{M}=-\frac{GM}{r^2}-\frac{GM}{r_e^2}\cdot\frac{r}{r_s}e^{-\frac{r}{r_e}}\, .
\label{effectivea}
\ee
At sufficiently large distances, the second term may become dominant and provides a linear acceleration towards the source.

As in the general relativity, one integrates Eq. (\ref{eq u}) and obtains the deflection angle of light $\alpha_\textmd{\small R}$ in a modified Rindler potential in Randers-Finslerian spacetime, to wit
\be
\label{deflection Rindler}
\alpha_\textmd{\small R}(r)=\frac{4GM}{r}f(r;r_s,r_e)\ ,
\ee
where
\be
f(r;r_s,r_e) \equiv 1-\frac{1}{2r_s}\int_{r}^{\infty} \frac{\frac{r^2}{r^{\prime2}}}{\sqrt{1-\frac{r^2}{r^{\prime2}}}}\frac{\left(2+\frac{r^2}{r^{\prime2}}\right)\left(1+\frac{r^\prime}{r_e}\right)e^{-\frac{r^\prime}{r_e}}-3\left(1+\frac{r}{r_e}\right)e^{-\frac{r}{r_e}}}{2\left(1-\frac{r^2}{r^{\prime2}}\right) }dr^{\prime}\, .
\label{f}
\ee
This integration can be computed numerically. The model parameters $r_s$ and the cutoff scale $r_e$ depend on the specific gravitational system and are to be determined by observations. For $r\gg r_e$, $\phi_M \rightarrow -\frac{GM}{r}$ and $\alpha_\textmd{\small R} \rightarrow \frac{4GM}{r}$. This is what we expect in general relativity and the Newtonian limit.

\section{Comparing with the Observations}
In this section, we use the modified gravity model to calculate the convergence $\kappa$ of the Bullet Cluster 1E0657-558. Before this, we first get the effective lens potential in the Randers-Finslerian spacetime (\ref{relativisticF}). We then use the potential to calculate the convergence.
\subsection{Effective Lens Potential}
We take a ``leap'' here. We do not deduce but give the the effective lens potential $\bar{\psi}$ in the Randers-Finslerian spacetime that will generate the deflection angle (\ref{deflection Rindler}). Then we use $\bar{\psi}$ to calculate the corresponding convergence $\kappa$. Hereafter, we use natural units in calculations, i.e. setting the speed of light $c=1$.

Einstein's general relativity predicts that a light ray passing by a spherical body of mass $M$ at a minimum distance $\xi$ is deflected by the angle
\be
\alpha =\frac{4GM}{\xi}\ ,~~~~~\xi\equiv\sqrt{x^2+y^2}\, .
\label{Einstein angle1}
\ee
The mass of the lens $M$ can be given as
\be
\label{mass}
M(\xi)=2\pi \int_0^\xi \Sigma(\xi^\prime) \xi^{\prime} d \xi^\prime\ ,
\ee
where $\Sigma(\xi^\prime)$ is the surface mass density distribution. It results from projecting the volume mass distribution of the ``lens'' $\rho(r)$ onto the lens plane (i.e. the $(x,y)$-plane) which is orthogonal to the line-of-sight direction (i.e. the $z$-direction) of the observer, to wit
\be
\Sigma(\xi)=\int_{-z_{\textmd{\small out}}}^{z_{\textmd{\small out}}} \rho(r)dz \ ,
\label{z out}
\ee
where $z \equiv \sqrt{r^2-x^2-y^2} = \sqrt{ r^2-\xi^2 }$ and $z_{\textmd{\small out}}\equiv\sqrt{ r_{\textmd{\small out}}^2-\xi^2 }$. $r_{\textmd{\small out}}$ denotes the outer radial extent of the galaxy cluster, which is defined as when $\rho$ drops to $\rho(r_{\textmd{\small out}})\simeq10^{-28}~\textmd{g}/\textmd{cm}^3$.

The ``Einstein angle'' (\ref{Einstein angle1}) can be rewritten in a vector form as \cite{Schineider1992}
\be
\hat{\alpha}=4G\int_{\textmd{\small R}^2}  d^2\vec{\xi}^\prime \Sigma(\vec{\xi}^\prime) \frac{\vec{\xi}-\vec{\xi}^\prime}{~|\vec{\xi}-\vec{\xi}^\prime|^2}\ ,
\label{Einstein angle3}
\ee
where
\be
d^2\vec{\xi}^\prime=\int_0^{2\pi} d\varphi \int_0^\xi d\xi^{\prime} \vec{\xi}^{\prime}
\ee
is the surface element of the lens plane.

With $\vec{\theta}=\frac{\vec{\xi}}{D_{\textmd{\small L}}}$, one can easily check that (\ref{Einstein angle3}) satisfies \footnote{
In the two-dimensional polar coordinates, $\nabla_{\vec{\xi}}\equiv\frac{\partial~}{\partial \vec{\xi}}=\hat{\mathbf{e}}_\xi\frac{\partial~}{\partial \xi}$~.}
 (see Section 4.1 in \cite{Peacock2003})
\be
\hat{\alpha} = \frac{D_\textmd{\small S}}{D_{\textmd{\small{LS}}}}\nabla_\theta \psi(\vec{\theta}) = \frac{D_\textmd{\small S}D_\textmd{\small L}}{D_{\textmd{\small {LS}}}}\nabla_{\vec{\xi}}~\psi(\vec{\xi})\ ,
\label{alpha and potential}
\ee
where
\be
\psi(\vec{\xi})=\frac{1}{\pi \Sigma_{\textmd{c}}} \int_{\textmd{\small R}^2} \Sigma(\vec{\xi}^\prime)~\textmd{ln}|\vec{\xi}-\vec{\xi}^\prime |~d^2\vec{\xi}^\prime\ ,~~~~~\Sigma_{\textmd{c}}\equiv \frac{D_{\textmd{\small S}}}{4\pi GD_\textmd{\small L} D_{\textmd{\small{LS}}}}\, .
\ee
$\Sigma_{\textmd{c}}$ is the critical surface density of the lens. $D_{\textmd{\small S}}$ is the angular distance between the observer and the source galaxy, i.e. the background. $D_{\textmd{\small L}}$ is the angular distance between the observer and the lens, i.e. the Bullet Cluster 1E0657-558, and $D_{\textmd{\small LS}}$ denotes the angular distance between the lens and the source galaxy.
The lens potential $\psi(\vec{\xi})$ obeys the two-dimensional Poisson's equation \footnote{In general, the Laplacian $\Delta$ in polar coordinates is given as
\be
\Delta \equiv \frac{1}{\xi}\frac{\partial~}{\partial\xi}\left(\xi \frac{\partial~}{\partial \xi}\right)+ \frac{1}{\xi^2}\frac{\partial^2 ~}{\partial\varphi^2}\ .\nonumber
\ee
For a $\varphi$-independent $\psi(\vec{\xi})$, one has $\frac{\partial\psi(\vec{\xi})}{\partial \varphi}=0$, and
\be
\Delta \psi \equiv \frac{1}{\xi}\frac{\partial~}{\partial\xi}\left(\xi \frac{\partial \psi}{\partial \xi}\right)\ .\nonumber
\ee
}
\be
\Delta \psi\equiv \nabla^2 \psi= 2\frac{\Sigma}{\Sigma_{\textmd{c}}}\ ,~~~~\Delta \equiv \frac{1}{\xi}\frac{\partial~}{\partial\xi}\left(\xi \frac{\partial~}{\partial \xi}\right)\ ,
\label{poisson equation}
\ee
In astronomy and astrophysics, the quantity $\frac{\Sigma}{\Sigma_{\textmd{c}}}$ in Eq. (\ref{poisson equation}) is defined as the convergence $\kappa$, which is also called the scaled surface mass density, i.e.
\be
\kappa \equiv \frac{\Sigma}{\Sigma_{\textmd{c}}}\, .
\label{kappa}
\ee

Consider a lens potential
\be
\label{gravitational potential}
\bar{\psi}(\vec{\xi})\equiv \psi(\vec{\xi})f(\vec{\xi};r_{s},r_e) = \frac{1}{\pi \Sigma_{\textmd{c}}} \int_{\textmd{\small R}^2} \Sigma(\vec{\xi}^\prime)f(\vec{\xi};r_{s},r_e)~\textmd{ln}|\vec{\xi}-\vec{\xi}^\prime |~d^2\vec{\xi}^\prime \ ,
\ee
where
\be
f(\vec{\xi};r_{s},r_e)\equiv \int_{-z_{\textmd{\small out}}}^{z_{\textmd{\small out}}} f(r;r_{s},r_e)dz
\label{f2d}
\ee
and $f(r;r_{s},r_e)$ is given by (\ref{f}). For the inner of the lens system, we have $\xi=\xi^\prime$. Thus, the potential (\ref{gravitational potential}) can be rewritten as
\be
\bar{\psi}(\vec{\xi})&=& \frac{1}{\pi \Sigma_{\textmd{c}}} \int_{\textmd{\small R}^2} \Sigma(\vec{\xi}^\prime)f(\vec{\xi}^\prime;r_{s},r_e)~\textmd{ln}|\vec{\xi}-\vec{\xi}^\prime |~d^2\vec{\xi}^\prime \\
&\equiv&\frac{1}{\pi \Sigma_{\textmd{c}}} \int_{\textmd{\small R}^2} \bar{\Sigma}(\vec{\xi}^\prime)~\textmd{ln}|\vec{\xi}-\vec{\xi}^\prime |~d^2\vec{\xi}^\prime \, .
\label{gravitational potential2}
\ee

Given the potential (\ref{gravitational potential2}) and using Eq. (\ref{alpha and potential}), one can reproduce the deflection angle $\alpha_\textmd{\small R}$ in the model, i.e.
\be
\label{deflection Rindler2}
\alpha_\textmd{\small R}(\xi)= \frac{4GM}{\xi}f(\xi;r_{s},r_0)\, .
\ee

\subsection{The $\Sigma$- and $\kappa$-Map of Bullet Cluster 1E0657-558}
\begin{figure*}
\begin{center}
\subfigure[~\textsf{$\Sigma$-Map}] {\label{fig:a}\scalebox{0.45}{\includegraphics{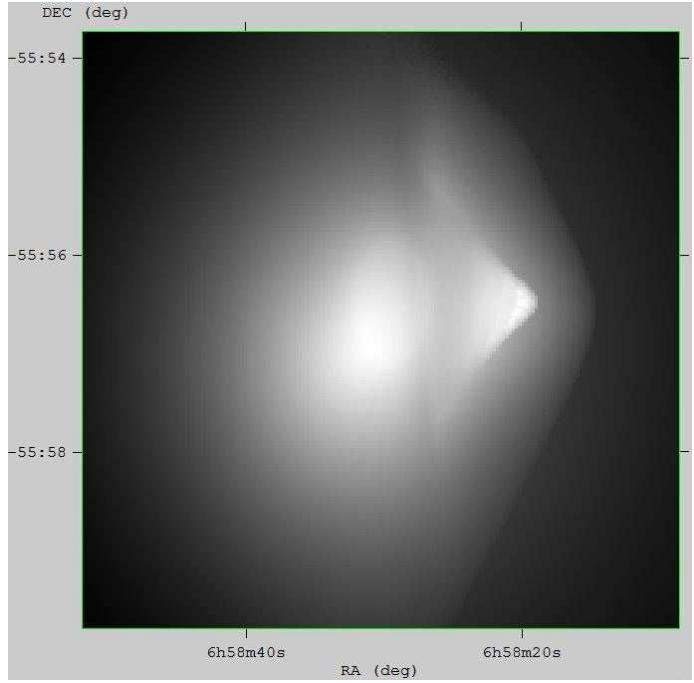}}}
\subfigure[~\textsf{Section of $\Sigma$-Map}] {\label{fig:b}\scalebox{0.45}{\includegraphics{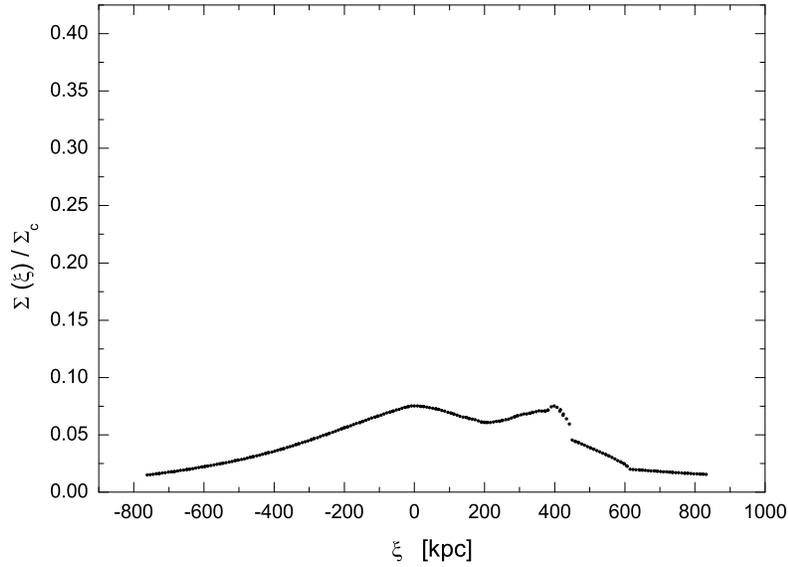}}}
\caption{The $\Sigma$-map from X-ray imaging observations of the Bullet Cluster 1E0657-558, November 15, 2006 data
release. (a)~The entire $\Sigma$-map is presented in the equatorial coordinate system J2000. DEC in the $y$-axis is short for ``Declination'' and the RA in the $x$-axis is short for ``Right Ascension''. The bright shockwave region at the right half of the map is the ICM gas of the subcluster. The main cluster gas locates at the brightly glowing region to the left of the subcluster gas. The released $\Sigma$-map has $185\times185$ pixels and a resolution of $8.5$ kpc/pixel. (b)~A subset of the $\Sigma$-map on a straight-line connecting the peak of the main cluster to that of the subcluster. The peak of the main cluster is taken to be the  referential center of the system, i.e. $\xi=0$ . The peak of the subcluster is located at $\xi\simeq398$ kpc.}
\label{fig1}
\end{center}
\end{figure*}

To calculate the convergence $\kappa$, one needs the surface mass density distribution $\Sigma(\xi)$ of the specific system. The $\Sigma$-map reconstructed from X-ray imaging observations of the Bullet Cluster 1E0657-558 is shown in Figure \ref{fig1}a. There are two distinct glowing peaks in Figure \ref{fig1}a -- the left one of the main cluster and the right one of the subcluster. A subset of the $\Sigma$-map on a straight-line connecting the peak of the main cluster to that of the subcluster is shown in Figure \ref{fig1}b.

For the Bullet Cluster system, the volume mass distribution of the ICM gas of the main cluster $\rho(r)$ is phenomenologically described by the King $\beta$-model \cite{Chandrasekhar1960,King1966,Cavaliere1976}
\be
\rho (r)=\rho_0\left[1+\left(\frac{r}{r_c}\right)^2 \right]^{-3\beta/2}\ ,~~~~r=\sqrt{x^2+y^2+z^2}\equiv\sqrt{\xi^2+z^2}\ ,
\label{King beta}
\ee
where the parameters $\rho_0$, $r_c$ and $\beta$ are determined to be \cite{Brownstein2007}
\be
\label{rho 0}
\rho_0 &=& 3.34 \times 10^5~~M_{\odot}/\textmd{kpc}^3\ , \\
\label{beta}
\beta &=& 0.803 \pm 0.013\ , \\
\label{r c}
r_c &=& 278.0 \pm 6.8~~\textmd{kpc}\, .
\ee
$M_{\odot}$ denotes the mass of the sun.

The outer radial extent of the Bullet Cluster system is given as
\be
r_{\textmd{\small out}}=r_c\left[\left(\frac{\rho_0}{10^{-28}~\textmd{g}/\textmd{cm}^3}\right)^{-2/3\beta}-1 \right]^{1/2}\simeq 2620~~\textmd{kpc}\, .
\label{r out2}
\ee
The radius of the main cluster is $\sim 1000$ kpc, thus we have $\xi=\xi^\prime$. The potential (\ref{gravitational potential2}) now becomes
\be
\label{gravitational potential3}
\bar{\psi}(\vec{\xi})= \frac{1}{\pi \Sigma_{\textmd{c}}} \int_{\textmd{\small R}^2} \bar{\Sigma}(\vec{\xi}^\prime)~\textmd{ln}|\vec{\xi}-\vec{\xi}^\prime |~d^2\vec{\xi}^\prime \ ,
\label{sigma bar}
\ee
where the effective surface mass density $\bar{\Sigma}(\xi)$ is defined as
\be
\label{sigma bar2}
\bar{\Sigma}(\xi) &\equiv& \int_{-z_{\textmd{\small out}}}^{z_{\textmd{\small out}}} \rho(r)f(r;r_{s},r_e)dz\ ,~~~~~z_{\textmd{\small out}}=\sqrt{ r_{\textmd{\small out}}^2-\xi^2 }=\sqrt{ 2620^2-\xi^2}~~~\textmd{kpc} \, .
\ee

Making use of (\ref{f}), (\ref{kappa}), ({\ref{King beta}) and (\ref{sigma bar2}), one finally obtains the convergence $\kappa$-map of the Bullet Cluster system
\be
\label{kappa2}
\bar{\kappa}(\xi) \equiv \frac{\bar{\Sigma}(\xi)}{\Sigma_{\textmd{c}}} &=& \frac{\rho_0}{\Sigma_{\textmd{c}}}\int_{-z_{\textmd{\small out}}}^{z_{\textmd{\small out}}} \left[1+\left(\frac{r}{r_c}\right)^2 \right]^{-3\beta/2}f(r;r_{s},r_e)dz\ ,
\ee
where
\be
f(r;r_s,r_e) \equiv 1-\frac{1}{2r_s}\int_{r}^{\infty} \frac{\frac{r^2}{r^{\prime2}}}{\sqrt{1-\frac{r^2}{r^{\prime2}}}}\frac{\left(2+\frac{r^2}{r^{\prime2}}\right)\left(1+\frac{r^\prime}{r_e}\right)e^{-\frac{r^\prime}{r_e}}-3\left(1+\frac{r}{r_e}\right)e^{-\frac{r}{r_e}}}{2\left(1-\frac{r^2}{r^{\prime2}}\right) }dr^{\prime}
\label{f2}
\ee
and
\begin{itemlist}
\item the parameters $\rho_0$, $r_c$ and $\beta$ are given in (\ref{rho 0}) to (\ref{r c}),
\item $z = \sqrt{r^2-x^2-y^2} \equiv \sqrt{ r^2-\xi^2 }$ and $z_{\textmd{\small out}}$ is given by (\ref{sigma bar2}),
\item for the  Bullet Cluster 1E0657-558, one has $\frac{D_\textmd{\small L} D_{\textmd{\small LS}}}{D_{\textmd{\small S}}}\simeq 540$ kpc. So $\Sigma_{\textmd{c}}$ in (\ref{kappa2}) takes a value of
\be
\Sigma_{\textmd{c}}\equiv \frac{D_{\textmd{\small S}}}{4\pi GD_\textmd{\small L} D_{\textmd{\small LS}}}\simeq 3.1 \times 10^9~~M_{\odot}/\textmd{kpc}^2\ ,
\ee
\item $r_s$ and $r_e$ are model parameters to be determined by fitting (\ref{kappa2}) to the $\kappa$-map reconstructed from the gravitational lensing survey.
\end{itemlist}

\begin{figure}
\begin{center}
\subfigure[~\textsf{$\kappa$-Map}] { \label{fig:a}
\scalebox{0.45}[0.45]{\includegraphics{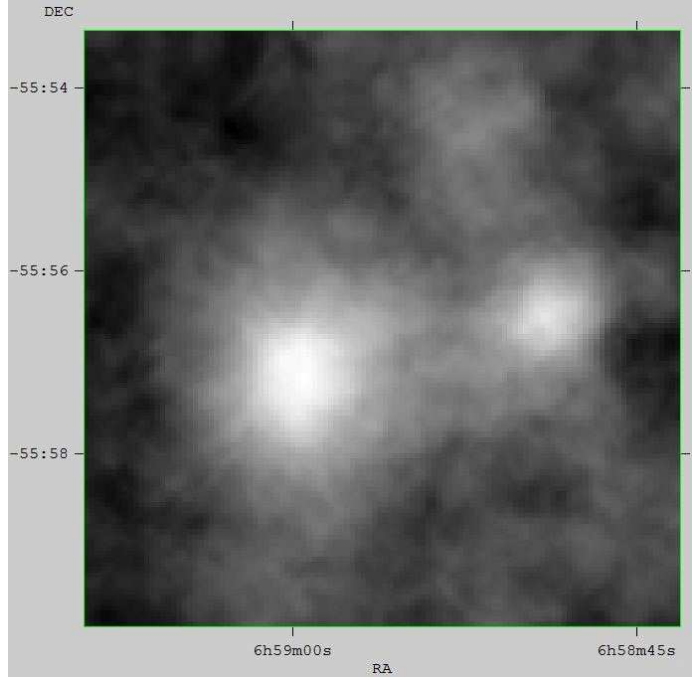}}
}
\subfigure[~\textsf{Section of $\kappa$-Map}] { \label{fig:b}
\scalebox{0.45}[0.45]{\includegraphics{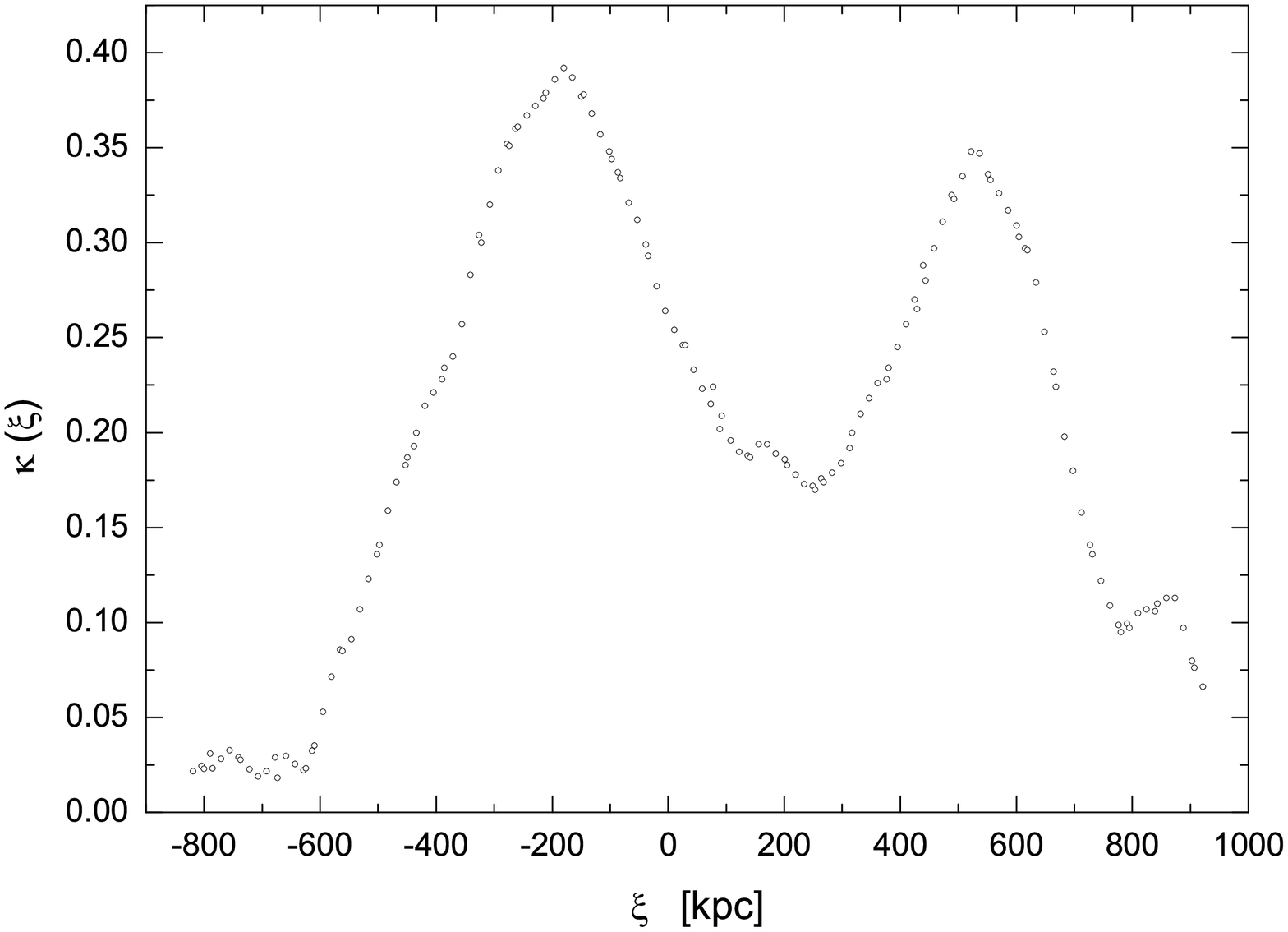}}
}
\caption{The $\kappa$-map reconstructed from the strong and weak gravitational lensing survey of the Bullet Cluster 1E0657-558, November 15, 2006 data
release. (a)~ The entire $\kappa$-map is presented in the equatorial coordinate system J2000. DEC in the $y$-axis is short for ``Declination'' and the RA in the $x$-axis is short for ``Right Ascension''. The bright blurred region at the left half of the map illuminates the convergence of the main cluster, while the smaller glowing one to the left corresponds to that of the subcluster. The released $\kappa$-map has $110 \times 110$ pixels and a resolution of $15.4$ kpc/pixel. (b)~A section of the $\kappa$-map on a straight-line connecting the peak of the main cluster to that of the subcluster. The peak of the main cluster is located at $\xi \simeq -180$ kpc and that of the subcluster is located at $\xi\simeq 522$ kpc. The $\xi=0$ point is chosen to be the same with that of the $\Sigma$-map in Figure \ref{fig2}b.}
\label{fig2}
\end{center}
\end{figure}

The $\kappa$-map obtained from the strong and weak gravitational lensing survey of the Bullet Cluster 1E0657-558 is presented in Figure \ref{fig2}a. One can see that the two distinct glowing regions in Figure \ref{fig2}a -- the left one of the main cluster and the right one of the subcluster -- somewhat depart from those shown in Figure \ref{fig1}a. A subset of the $\kappa$-map on a straight-line connecting the peak of the main cluster to that of the subcluster is also shown in Figure \ref{fig2}b.

A section of the $\bar{\kappa}$-map (\ref{kappa2}), which crossing the two peaks is plotted in Figure \ref{fig3}. For a qualitative illustration, different values of parameter set $(r_s,r_e)$ are plotted for comparison instead of carrying a best-fit. \textbf{The ``best-fit'' values of parameters $r_s$ and $r_e$ with a $5\%$ error\footnote{\textbf{Variation of $r_s$ and $r_e$ from their ``best-fit'' values leads to a deviation of $\Delta M/M_K$ and $\kappa$ from their extremal points (see Table \ref{ta1}). We consider both of these deviations of $\Delta M/M_K$ and $\kappa$ within a $5\%$ level to obtain the corresponding confidence regions of $r_s$ and $r_e$. Gaussian prior distributions of the parameters are assumed.}} are presented in Table {\ref{ta1}}.} Plot for $f(r;r_s,r_e)$ is presented in Figure \ref{fig6}(a).
Our approach follows a sequence of approximations:
\begin{itemize}
\item Take the main cluster thermal profile to be isothermal.
\item Neglect the subcluster for zeroth order approximation.
\item Perform the fit using a section of the $\kappa$-map on a straight-line connecting the peak of the main cluster to that of the subcluster and then extrapolating it to the entire map.
\item Take the $\Sigma$-peak of the main cluster as the center of the gravitational system, and project the section of the $\kappa$-map onto that of the $\Sigma$-map to make the two overlay for comparison.
\end{itemize}

\begin{figure}
\begin{center}
\scalebox{0.6}[0.6]{\includegraphics{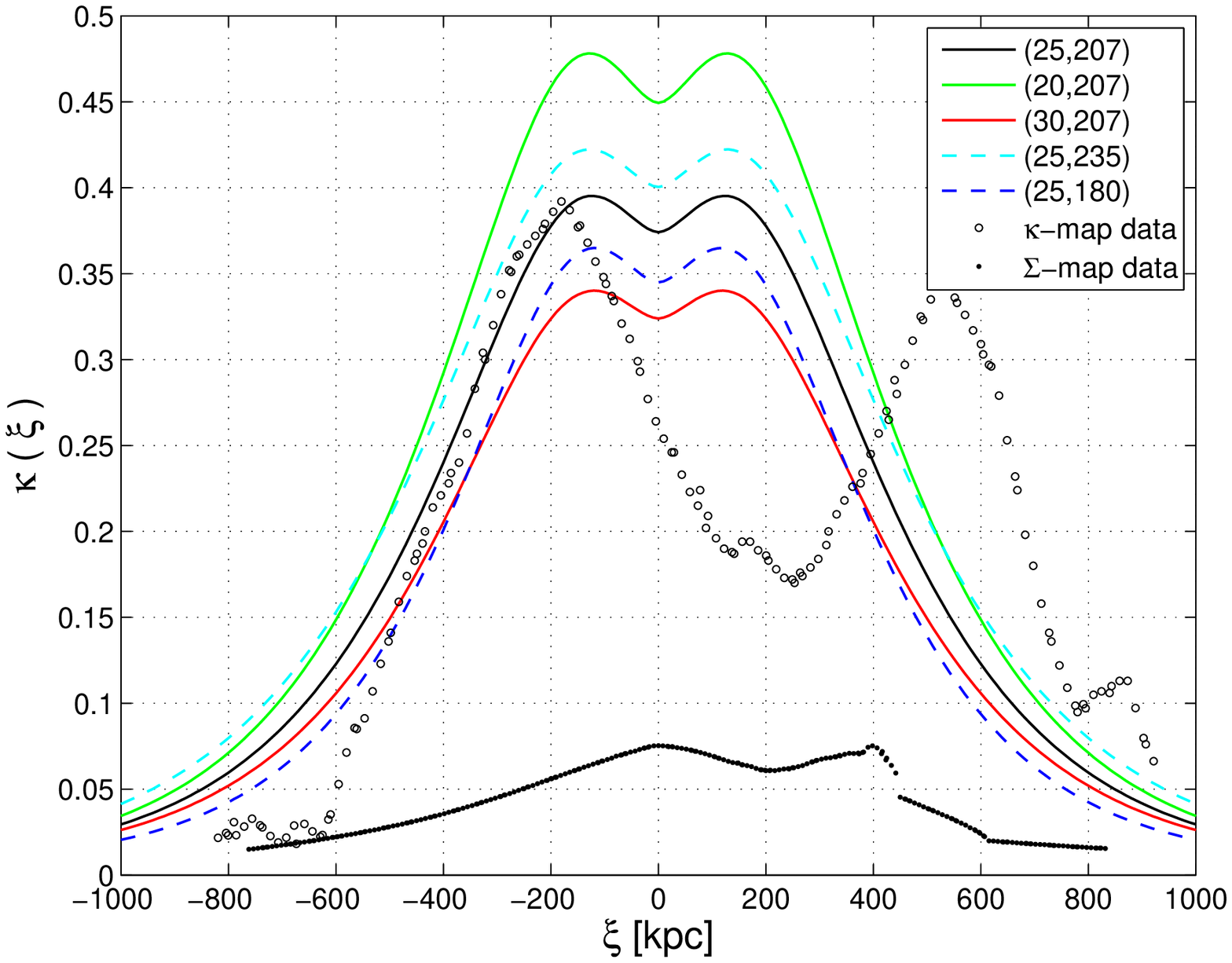}}
\caption{Cross sections of the model-predicted $\bar{\kappa}$-map and the $\Sigma$-, $\kappa$-map reconstructed from the November 15, 2006 data release. The solid and dashed lines denote the sections of the $\bar{\kappa}$-map (\ref{kappa2}) predicted by the Randers-Finslerian model with a modified Rindler potential (\ref{modified potential}) for parameters ($r_s$, $r_e$) listed in Table \ref{ta1}. The sections of the $\Sigma$- and $\kappa$-map obtained by observations are respectively represented by small black dots and circles as in Figure \ref{fig1}b and \ref{fig2}b.}
\label{fig3}
\end{center}
\end{figure}

\subsection{The Isothermal Temperature Profile}
Besides the convergence $\kappa$, the surface temperature $T$ of the cluster obtained from the X-ray spectrum analysis is also another observed quantity which should be used to constrain a model. Assuming an isotropic and isothermal gas profile with temperature $T$, one can calculate dynamical mass $M_\textmd{T}$ of the main cluster as a function of the radial position $r$ and the temperature $T$. By comparing it with the result given by integrating the King $\beta$-model (\ref{King beta}), we obtain a more rigorous constraint on the model parameters $r_s$ and $r_e$.

The collisionless Boltzmann equation of a spherical system in hydrostatic equilibrium reads
\be
\frac{d}{dr}(\rho(r) \sigma_{r}^{2}) + \frac{2\rho(r)}{r}\left(\sigma_{r}^{2} - \sigma_{\theta,\phi}^{2}\right) =
-\rho(r) \frac{d\Phi(r)}{dr}\ ,
\label{CBE}
\ee
where $\Phi(r)$ is the gravitational potential of the system and $\sigma_{r}$ and $\sigma_{\theta,\phi}$ are respectively the mass-weighted
velocity dispersions in the radial and ($\theta, \phi$) directions. Given an isotropic gas sphere distribution $\rho(r)$ with a temperature profile $T(r)$, one has
\be
\sigma_{r}^2 = \sigma_{\theta,\phi}^2 = \frac{k_B T(r)}{\mu_{A} m_{p}}\ ,
\label{isotropic}
\ee
where $k_B$ is Boltzmann's constant, $\mu_{A} \simeq 0.609$ is the mean atomic weight and $m_{p}$ is the proton mass.
Eq. (\ref{CBE}) becomes
\be
\frac {d}{dr}\left(\frac{k_B T(r)}{\mu_{A} m_{p}} \rho(r)\right) = -\rho(r) \frac{d\Phi(r)}{dr}\, .
\label{isotropicCBE}
\ee
For the main cluster of the Bullet Cluster system, the ICM gas distribution $\rho(r)$ is fit by an isotropic and isothermal King $\beta$-model (\ref{King beta}) with the temperature $T(r)=T$. Solving Eq. (\ref{isotropicCBE}) for the gravitational acceleration, one obtains
\be
\nonumber
a(r) \equiv  - \frac{d\Phi(r)}{dr} &=&  \frac{k_B T}{\mu_{A} m_{p} r} \left[ \frac{d \ln(\rho(r))}{d \ln(r)} \right] \\
&=&  -\frac{3\beta k_B T}{\mu_{A} m_{p}} \left(\frac{r}{r^{2}+r_{c}^{2}}\right)\, .
\label{accelerationProfile}
\ee
Replacing $a(r)$ in (\ref{accelerationProfile}) with the effective acceleration $a_M$ in (\ref{effectivea}), to wit
\be
a(r)= a_{M}(r)=-\frac{GM_T}{r^2}\left(1+\frac{r^3}{r_e^2 r_s}e^{-\frac{r}{r_e}}\right)\ ,
\ee
we obtain the relation between the dynamical mass $M_\textmd{T}$ as a function of the radial position $r$ and the temperature $T$, to wit
\be
M_\textmd{T}(r) = \frac{3\beta k_B T}{\mu_{A} m_{p}G} \left(\frac{r^3}{r^{2}+r_{c}^{2}}\right)\cdot\left(1+\frac{r^3}{r_e^2 r_s}e^{-\frac{r}{r_e}}\right)^{-1}\, .
\label{temperaturemass}
\ee

On the other hand, the mass profile of the main cluster is given by the King $\beta$-model as
\be
M_\textmd{K}(r) &=& 4\pi \int_0^r \rho(r^\prime)r^{\prime 2}dr^{\prime}\nonumber \\
&=& 4\pi\rho_0 \int_0^r \left[1+\left(\frac{r^\prime}{r_c}\right)^2 \right]^{-3\beta/2}r^{\prime 2}dr^{\prime}\, .
\label{King mass}
\ee
The detection in X-ray by the Einstein IPC, ROSAT and ASCA observations constrained the temperature of the main cluster to be $T=17.4\pm 2.5$ keV (with $12.3\%$ error) \cite{Tuck1998} and $T=14.5_{-2.0}^{+1.7}$ keV (with $6.5\%$ error) \cite{Liang2000}. It was later reported by Markevitch \cite{Markevitch2002} that $T=14.8_{-1.2}^{+1.7}$ keV (with $4.5\%$ error). Fixing the temperature $T$ in (\ref{temperaturemass}) to be the observed center value $T=14.8$ keV and by comparing the two $M(r)$ in (\ref{King mass}) and (\ref{temperaturemass}) at the radial distance $r=1000$ kpc (which is also the boundary of the reconstructed $\kappa$- and $\Sigma$-map), one can put a constraint on the model parameters $r_s$ and $r_e$. The results are presented in Table \ref{ta1} and Figure \ref{fig3}.

\begin{table}[ph]
\tbl{Mass discrepancies of different parameter set $(r_s, r_e)$. The isothermal temperature of main cluster is fixed to be $T=14.8$ keV as reported by Markevitch. `$\Delta M$' represents the mass difference between (\ref{temperaturemass}) and (\ref{King mass}), i.e. $\Delta m\equiv|M_T-M_K|$. The last column presents the peak values of the $\kappa$-map given by (\ref{kappa2}) at $r\sim 180$ kpc. The ``best-fit'' result of parameters $(r_s, r_e)$ are highlighted in boldface in the second row. The first and third row show that a variation of $r_s$ near $(r_s, r_e)=(25,207)$ will leads to a bad $\Delta M$ and $kappa$. The fourth and fifth row show the same result for a variation of $r_e$ near $(r_s, r_e)=(25,207)$. \textbf{The errors of the ``best-fit'' result are given by considering a $5\%$ deviation of both $\Delta M/M_K$ and $\kappa$ from their extremal values.}}
{\begin{tabular}{@{}ccccc@{}} \toprule
$T$ & $r_s$ & $r_e$ & $\Delta M/M_K$ & $\kappa$ \\
 (keV) & (kpc) & (kpc) & (\%) & (peak values)\\ \colrule
14.8 & 20 & 207 & 21.37 & 0.48\\
\textbf{14.8} & \textbf{25$\pm$2.40} & \textbf{207$\pm$11.15} & \textbf{0.01} & \textbf{0.38}\\
14.8 & 30 & 207 & 15.16 & 0.34\\
14.8 & 25 & 180 & 32.05 & 0.37\\
14.8 & 25 & 235 & 32.74 & 0.43\\\botrule
\end{tabular} \label{ta1}}
\end{table}

\subsection{A Randers Plus Dark Matter Model}
Stavrinos \textit{et al.}'s work \cite{Stavrinos2008} showed that the Randers-type spacetime does not forbid the existence of dark matter in cosmology. Thus it would be interesting to consider dark matter in the Randers-Finslerian spacetime. We consider the most popular Navarro-Frenk-White(NFW) profile of the dark matter \cite{Navarro1996,Navarro1997}. The mass density in (\ref{sigma bar2}) is now given as
\be
\rho (r)= \rho_\textmd{K}(r) + \rho_{\textmd{DM}}(r)\ ,
\label{new rho}
\ee
where
\be
\rho_{\textmd{DM}}(r)=\frac{\rho_d r_d^3}{r^3+r_d^3}\, .
\label{NFW}
\ee
$\rho_d$ is the central dark matter density and $r_d$ is the core radius. Now the convergence $\bar{\kappa}$ is given as
\be
\label{kappa3}
\bar{\kappa}(\xi) \equiv \frac{\bar{\Sigma}(\xi)}{\Sigma_{\textmd{c}}} &=& \frac{1}{\Sigma_{\textmd{c}}}\int_{-z_{\textmd{\small out}}}^{z_{\textmd{\small out}}} \left[\rho_\textmd{K}(r)+\rho_{\textmd{DM}}(r) \right]f(r;r_{s},r_e)dz\ ,
\ee
where $\rho_ \textmd{K}(r)$ is given by (\ref{King beta}) and $\rho_{\textmd{DM}}(r)$ is given by (\ref{NFW}). From WMAP's seven-year result \cite{Jarosik2011}, we know that the total amount of matter (or energy) in the universe in the form of dark energy about $73\%$ and dark matter about $23\%$~. This leaves the ratio of baryonic matter at only $\sim4\%$. The ratio can be parameterized as $\eta\equiv M_{\textmd{DM}}/M_{\textmd{b}}$, where $M_{\textmd{DM}}$ denotes the total volume mass of dark matter in a region and $M_{\textmd{b}}$ refers to that for ordinary baryonic matter. In this paper, we fix this ratio to be $\eta= 6$. Given (\ref{NFW}) and (\ref{King beta}), one can integrate to get $\rho_{d}$ as a function of $\rho_0$ and $\eta$, i.e. $\rho_{d}=\rho_d(\rho_0, \eta; r_d)$, leaving $r_d$ the only free parameter in the NFW profile in our model. The convergence $\bar{\kappa}$ and the mass profile of the main cluster are now given as
\be
\bar{\kappa}(\xi) \equiv \frac{\bar{\Sigma}(\xi)}{\Sigma_{\textmd{c}}} &=& \frac{1}{\Sigma_{\textmd{c}}}\int_{-z_{\textmd{\small out}}}^{z_{\textmd{\small out}}} \left[\rho_\textmd{K}(r)+\rho_{\textmd{DM}}(r; r_d, \eta) \right]f(r;r_{s},r_e)dz\ ,
\label{kappa plus}
\ee
and
\be
M_\textmd{K}(r) &=& 4\pi \int_0^r \left(\rho_\textmd{K}(r^\prime)+\rho_{\textmd{DM}}(r^\prime;r_d,\eta)\right) r^{\prime 2}dr^{\prime}\, .
\label{King mass plus}
\ee
Thus for the Randers$+$dark matter model, we have three free parameters $r_s$, $r_e$ and $r_d$. The numerical results are given in Figure \ref{fig4} and Table \ref{ta2}.

In Table \ref{ta2}, the first three rows show that we take a declining journey of $r_d$ to get a less mass discrepancy $\Delta M$ at the cost of a rapidly rising $\kappa$. (A small $r_d$ means a more condense dark matter core and a more sparse outskirt for the NFW profile.) Such a result means that we have added too much dark matter into the core of the main cluster thus the $\kappa$ flies. Then in the fourth row, we strip out the Finslerian effect, leaving only the dark matter and the baryonic matter, by setting $(r_d, r_s, r_e)=(440, 1000, 2)$ (large $r_s$ and small $r_e$ will radically suppress the Finslerian effect at large distances, for in (\ref{lambda}) $\lambda \rightarrow 1$.) It still yields too large $\kappa$ ($\simeq 0.48$) compared to the observed value $\kappa\simeq 0.38$. This result implies that an averaged distribution density of cold dark matter in cosmological senses fails to reproduce the observed convergence $\kappa$ of the Bullet Cluster.

Instead we take another approach to fill up the mass discrepancy shown in the first row: we tune up $r_e$ to ``turn on'' the Finslerian effect to fill up the ``mass gap'' $\Delta M$ at the cluster center. But the results in the last two rows demonstrate that this way does not work too. For one to obtain a ideal $\Delta M$, the convergence $\kappa$ have greatly exceeded the observed value. Thus for a Randers$+$dark matter model, the mass discrepancy $\Delta M$ and the convergence $\kappa$ is like the two ends of a see saw. It can not both be lowered at the same time. One possible reason for this may be that a dark matter-to-baryonic matter ratio $\eta\simeq6$ is too large. Another sign of this is that at the center of the main cluster, the compound model fails to reproduce the gravitational potential offset from the mass center. The Finslerian effect seems to be overwhelmed by the dark matter background. Since the mass ratio of dark matter and its type are not the subjects of this paper, we will not discuss it here. $f(r; r_s, r_e)$ for different parameters are plotted in Figure \ref{fig6}.

\textbf{To compare with the Randers and Randers$+$dark matter models, we also plot the results for the concordance $\Lambda$-CDM cosmological model \cite{Jarosik2011}. This can be implemented by setting $r_s\rightarrow \infty$ or/and $r_e\rightarrow 0$ in the equation (\ref{kappa plus}). It will result in $f(r;r_s,r_e)=1$ and leave us the convergence $\kappa$ in a $\Lambda$-CDM model:}
\be
\bar{\kappa}(\xi) \equiv \frac{\bar{\Sigma}(\xi)}{\Sigma_{\textmd{c}}} &=& \frac{1}{\Sigma_{\textmd{c}}}\int_{-z_{\textmd{\small out}}}^{z_{\textmd{\small out}}} \left[\rho_\textmd{K}(r)+\rho_{\textmd{DM}}(r; r_d, \eta) \right] dz\ ,
\label{kappa plus2}
\ee
\textbf{Together with (\ref{King mass plus}), we give our numerical results in Table \ref{ta3}. Two comments should be given about the results: First, it fails to give a reasonable ($\leq 5\%$) mass discrepancy $\Delta M/M_\textmd{K}$ together with an observations-compatible convergence $\kappa$ (highlighted in boldface respectively in Table \ref{ta3}). Second, the $\Delta M/M_\textmd{K}$ and $\kappa$ we get for $(r_s, r_e)=(\infty, 0)$ are not so much different from those in the first and fourth row in Table \ref{ta2}. One reason for this is that setting $(r_s, r_e)=(1000, 2)$ is already enough for one to strip out the Finslerian impacts on the dynamical mass and the convergence $\kappa$. The other one is that at the center of the main cluster, the Finsler effects are ``drowned'' by the dark matter background with a dark matter-to-baryons mass ratio $\eta \sim6$, just like the case in the Randers+dark matter model. We plot the results of the $\Lambda$-CDM model in Figure {\ref{fig4}} for comparison.}

\begin{figure}
\begin{center}
\scalebox{0.52}[0.52]{\includegraphics{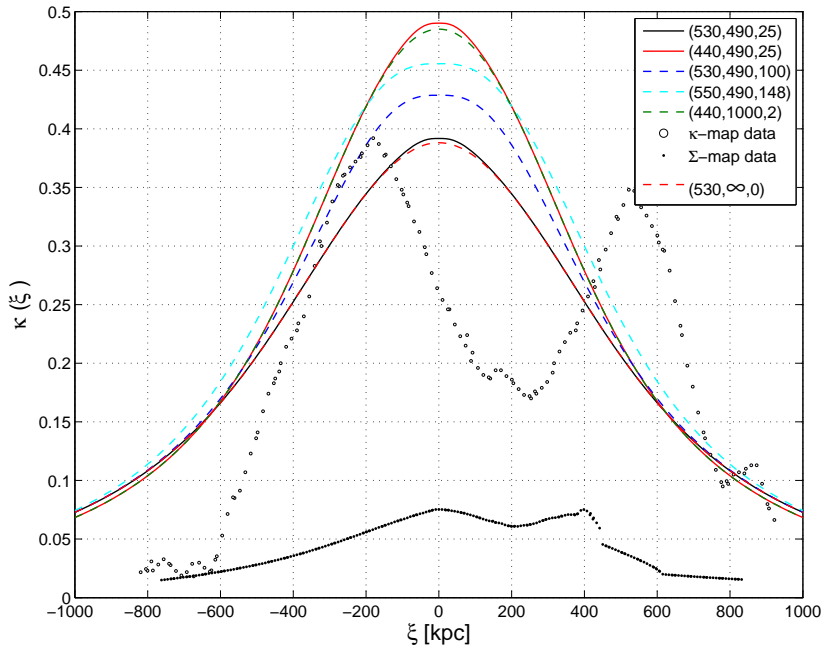}}
\caption{Cross sections of the model-predicted $\bar{\kappa}$-map and the $\Sigma$-, $\kappa$-map reconstructed from the November 15, 2006 data release. The solid and dashed lines except the bottom one denote the sections of the $\bar{\kappa}$-map (\ref{kappa plus}) predicted by the Randers$+$dark matter model with parameters ($r_d$, $r_s$, $r_e$) listed in Table \ref{ta2}. \textbf{The red dashed line represents the ``best-fit'' result for the $\Lambda$-CDM model (see Table \ref{ta3} and the discussions in the last paragraph of subsection 3.4).} The sections of the $\Sigma$- and $\kappa$-map obtained by observations are respectively represented by small black dots and circles as in Figure \ref{fig1}b and \ref{fig2}b.}
\label{fig4}
\end{center}
\end{figure}

\begin{table}[ph]
\tbl{Mass discrepancies of different parameter set $(r_d, r_s, r_e)$. The isothermal temperature of main cluster is fixed to be $T=14.8$ keV as reported by Markevitch. `$\eta$' is mass ratio between the baryonic matter and the non-baryonic dark matter
`$\Delta M$' represents the mass difference between (\ref{temperaturemass}) and (\ref{King mass plus}), i.e. $\Delta M=|M_\textmd{T}-M_\textmd{K}|$. The last column presents the peak values of the $\kappa$-map given by (\ref{kappa plus}) at $r\sim 180$ kpc. }
{\begin{tabular}{@{}cc|cccccc@{}} \toprule
$T$ & $\eta$ & $r_d$ & $r_s$ & $r_e$ & $\Delta M/M_\textmd{K}$ & $\kappa$ \\
 (keV) &  & (kpc) & (kpc) & (kpc) & (\%) & (peak values) \\ \colrule
\textbf{14.8} & \textbf{6} & \textbf{530} & \textbf{490} & \textbf{25} & \textbf{9.13} & \textbf{0.39} \\
14.8 & 6 & 470 & 490 & 25 & 3.08 & 0.42 \\
14.8 & 6 & 440 & 490 & 25 & 0.05 & 0.49 \\
14.8 & 6 & 440 & 1000 & 2 & 0.04 & 0.48 \\
14.8 & 6 & 530 & 490 & 100 & 8.29 & 0.43 \\
14.8 & 6 & 530 & 490 & 148 & 0.07 & 0.46 \\ \botrule
\end{tabular} \label{ta2}}
\end{table}

\begin{figure}
\begin{center}
\subfigure[~\textsf{Randers model only}] { \label{fig:a}
\scalebox{0.55}[0.55]{\includegraphics{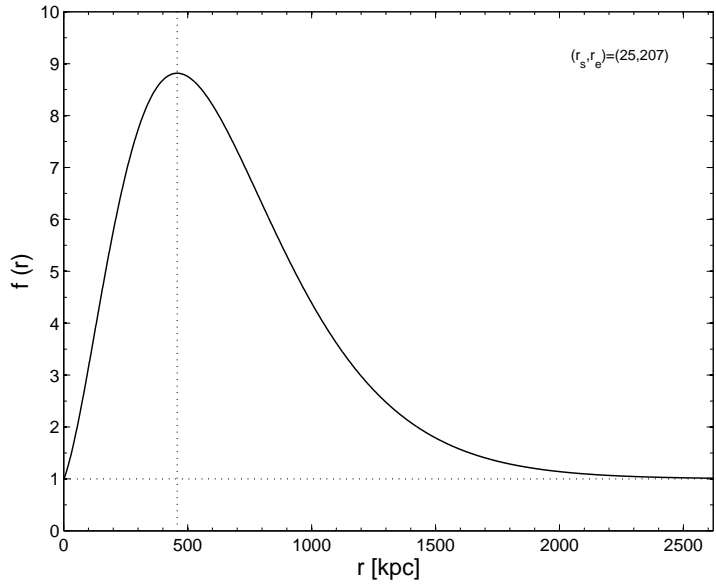}}
}
\subfigure[~\textsf{Randers model $+$ dark matter}] { \label{fig:b}
\scalebox{0.55}[0.55]{\includegraphics{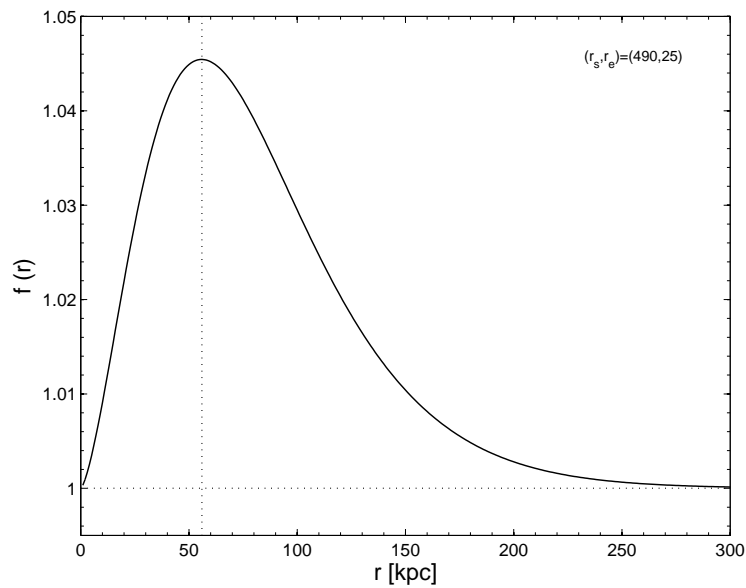}}
}
\caption{Plot for the dimensionless Finslerian factor $f(r)$ in Eq. (\ref{f}) vs. the radial distance $r$ in unit of kpc. ~(a)~is for Randers model without any dark matter. ~(b)~is for the Randers$+$dark matter model. The parameter values are the ``best-fit'' value which are presented in boldface in Table \ref{ta1} and Table \ref{ta2}.}
\label{fig6}
\end{center}
\end{figure}

\begin{table}[ph]
\tbl{Mass discrepancies for the $\Lambda$-CDM model. The isothermal temperature of main cluster is fixed to be $T=14.8$ keV as reported by Markevitch. `$\eta$' is mass ratio between the baryonic matter and the non-baryonic dark matter
`$\Delta M$' represents the mass difference between (\ref{temperaturemass}) and (\ref{King mass plus}), i.e. $\Delta M=|M_\textmd{T}-M_\textmd{K}|$. The last column presents the peak values of the $\kappa$-map given by (\ref{kappa plus2}) at $r\sim 180$ kpc. \textbf{Reasonable results are highlighted in boldface. The result in the first row is plotted in Figure \ref{fig4} for comparison.}}
{\begin{tabular}{@{}cc|cccccc@{}} \toprule
$T$ & $\eta$ & $r_d$ & $r_s$ & $r_e$ & $\Delta M/M_\textmd{K}$ & $\kappa$ \\
 (keV) &  & (kpc) & (kpc) & (kpc) & (\%) & (peak values) \\ \colrule
14.8 & 6 & 530 & $\infty$ & 0 & 9.25 & \textbf{0.38} \\
14.8 & 6 & 440 & $\infty$ & 0 & \textbf{0.04} & 0.48 \\ \botrule
\end{tabular} \label{ta3}}
\end{table}

\subsection{The Galactic Regime}
The specific form of $\lambda(r)$ in (\ref{lambda}) is postulated at cluster scales. It would be interesting to see its galactic-scale behaviors.
The potential (\ref{modified potential}) is given by solving the equation of motion (\ref{eq u}) which is derived from (\ref{relativisticF}) (see the Appendix). It recovers some features of the galactic rotation curves predicted by Grumiller's model, which was considered to be a good phenomenological fit to the observational data \cite{Grumiller2010}. From $v\sim\sqrt{ar}$ and (\ref{effectivea}), we obtain a new formula for the velocity profile of a galaxy:
\begin{equation}
\label{modified velocity}
v(r)\simeq\sqrt{\frac{GM}{r}+\frac{GM}{r_s}\left(\frac{r}{r_e}\right)^{2}e^{-\frac{r}{r_e}}}\, .
\end{equation}
To describe galaxies, we assume that the total mass $M\simeq10^{11} M_{\odot}$ (instead of $M\simeq10^{14} M_{\odot}$ for the Bullet Cluster system). For a qualitative illustration, the plot of profile (\ref{modified velocity}) for $r_s\simeq1$ kpc and the cutoff scale $r_e\simeq80$ kpc is shown in Figure \ref{fig6}. From the figure, we can see that our model in galactic limit yields an approximately flattened rotation curve of spiral galaxy. It is qualitatively consistent with the MOND and Grumiller's model. The velocity scale where the rotation curve flattens is $\sim 240$ km/s, which is in reasonable agreement with Grumiller's prediction and the observational data. A possible divergence of the velocity (\ref{modified velocity}) at large radial distances is reconciled by the exponential factor to yield a physical result.

\begin{figure}
\begin{center}
\includegraphics{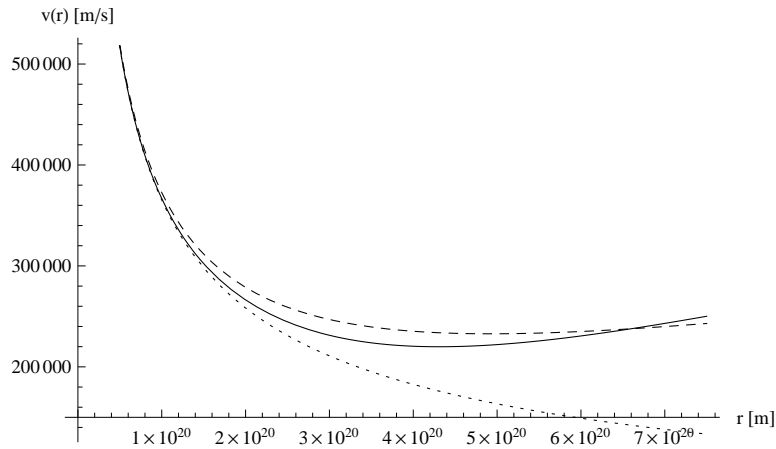}
\caption{Rotation curves of a spiral galaxy $v(r)$ vs. $r$ in unit m/s vs. m ($1$ kpc $\simeq 3\times 10^{19}$ m). The dashed line denotes the velocity profile predicted by Grumiller (which is qualitatively compatible with the MOND at the distance scale of several kpcs and considered a good phenomenological fit to the observational data). The solid line denotes the results of our model for the same total galactic mass. The dotted line which sinks into the bottom is given by Newton's theory, which fails to account for the observations.}
\label{fig6}
\end{center}
\end{figure}

\section{Conclusions and Discussions}
As a cluster-scale generalization of Grumiller's gravity model, we presented a gravity model in a navigation scenario in Finslerian geometry \cite{Bao2004,Zermelo1931}. The galactic limit of the model shared some qualitative features of Gumiller's result and the MOND. It yielded approximately the flatness of the rotational velocity profile at the radial distance of several kpcs. It also gave observations-compatible velocity scales for spiral galaxies at which the curves become flattened.

We also studied the gravitational deflection of light in such a framework and the deflection angle was obtained. The modified convergence $\kappa$ formula of a galaxy cluster showed that the peak of the gravitational potential has chances to lie on the outskirts of the baryonic mass center. For the Bullet Cluster 1E0657-558 system, the later refers to the center of the ICM gas profile of the main cluster. Taking the mass ratio between dark matter and baryonic matter $\eta$ to be a factor of 6 and assuming an isotropic and isothermal ICM gas profile with temperature $T=14.8$ keV (which is the center value given by Markevitch {\it et al.}'s observations \cite{Markevitch2002}), we used the collisionless Boltzmann equation to calculate the dynamical mass $M_\textmd{T}$ of the main cluster. We obtained a good match between $M_\textmd{T}$ and that given by King $\beta$-model and simultaneously ameliorated the shape of the convergence $\kappa$ curve. For comparison, we also consider a Randers$+$dark matter model. Numerical results showed that it fails to fill up the mass difference between $M_\textmd{T}$ and that given by King $\beta$-model. A smaller $\eta$ seems to be able to reconcile this dilemma. \textbf{Similar results were also obtained for the concordance $\Lambda$-CDM model.} More careful investigations are needed for drawing a confirmative conclusion.


A few comments should be given on the $\lambda$ in the action (\ref{relativisticF}). First, for a time-independent radial ``wind'' in the manifold, $\lambda$ is a function of $r$. Any $\lambda(r)$ that giving a small-enough $|\tilde{b}^r|$ would be considered valid in Finslerian geometry. But not all these mathematically valid $\lambda(r)$ would be acceptable for constructing a physical model. A both mathematically and physically valid $\lambda(r)$ should at least satisfy the following conditions: 1) $|\tilde{b}^r|=\sqrt{(1-\lambda(r))/h_{rr}}/\lambda(r)<1$, such that the positivity of $F$ holds; 2) experiments- and observations-compatibility. The $\lambda$ in (\ref{lambda}) satisfies both of these conditions. For $(r_s,r_e)=(25,207)$ and $(1,80)$ , $|\tilde{b}^r|\sim 10^{-13}\ll 1$.

Second, besides the mathematical validity of $\lambda(r)$ we chose, from Appendix one can see that if we redefine $\lambda$ as $\lambda = 1- \frac{GM}{r_s}(1+\frac{r}{r_e})e^{-\frac{r}{r_e}} \equiv 1+\phi_{\lambda}$, the non-vanishing component of the geodesic equation will give that the gravitational potential in Finslerian spacetime is $\phi_{M}\equiv \left(\phi_{N}+\phi_{\lambda}\right)$ and $\phi_{N}\equiv-\frac{GM}{r}$ is the Newtonian potential. It means that the results have a close relationship with $\lambda$. On the other hand, as a physical model, the specific form of $\lambda$ should be determined by the local spacetime symmetry, which cannot be deduced from the gravity theory. It is not the fruit but a prior stipulation of the theory. There is no physical principle or equation to constrain its form. Professor Shen's description of Finsler geometry (private conversation) may help us in understanding this --- ``Riemann geometry is `a white egg', for the tangent manifold at each point on the Riemannian manifold is isometric to a Minkowski spacetime. However, Finsler geometry is `a colorful egg', for the tangent manifolds at different points of the Finsler manifold are not isometric to each other in general.'' In physics, it implies that our nature does not always prefer an isotropic gravitational force. It is also ``colorful'', as we have seen in case of Bullet Cluster 1E0657-558.

Last but not the least, as a physical model at cluster scales, the $\lambda(r)$ should be subject to more observational tests, just like the NFW profile of the dark matter \cite{Navarro1996,Navarro1997}. A new challenge is posed by the Abell 520 cluster \cite{Jee 2012}. A combined constraint on the model should be carried out. Relevant research are currently undertaken. We hope that it would help to constrain the form of $\lambda$, which embodies the symmetry of Finsler spacetime.

\section*{Appendix}
By combining the non-vanishing components of the geodesic equations (\ref{geodesics}), one obtains the relation between the radial distant $r$ and the time $t$ \cite{Li2011},
\be
\label{newton limit}
\frac{AE^2}{B^2}\left(\frac{dr}{dt}\right)^2+\frac{J^2\lambda}{r^2}-\frac{E^2}{B}=-C\ , \nonumber
\ee
where $A(r)\equiv\lambda^{-2}\left(1-\frac{2GM}{r}\right)^{-1}$ and $B(r)\equiv\lambda^2\left(1-\frac{2GM}{r}\right)$. $E$ is an integration constant (see \cite{Li2011} for details). For photons, the constant $C=0$.
The above equation can be rewritten as
\be
A^3 \left(\frac{dr}{dt}\right)^2+\frac{J^2\lambda}{r^2 E^2}-\frac{1}{B} = 0\, .
\label{newton limit2}
\ee
In the Newtonian limit and the weak-field approximation, the quantities $\frac{J^2}{r^2}, \left(\frac{dr}{dt}\right)^2, E^2-1, \frac{GM}{r}$ are small. To first order of these quantities (remembering that the leading order terms of $A$ and $B$ are 1), Eq. (\ref{newton limit2}) becomes
\be
\left(\frac{dr}{dt}\right)^2+\frac{J^2}{r^2}-\frac{1}{B}  = 0\, .
\label{newton limit3}
\ee

Redefining $\lambda$ in (\ref{lambda}) as $\lambda = 1- \frac{GM}{r_s}(1+\frac{r}{r_e})e^{-\frac{r}{r_e}} \equiv 1+\phi_{\lambda}$, one has
\be
-\frac{1}{B} &\equiv& -\lambda^{-2} \frac{1}{1-\frac{2GM}{r}} \nonumber \\
&=& -\frac{1}{\left(1+\phi_{\lambda}\right)^2}\frac{1}{1-\frac{2GM}{r}}  \nonumber \\
&\simeq & -\left(1-2\phi_{\lambda}\right) \left( 1+\frac{2GM}{r}\right) \nonumber \\
&\simeq & -\left(1+2\frac{GM}{r}-2\phi_{\lambda}\right)\nonumber \\
&=& -1+2\phi_M\ ,
\label{B}
\ee
where $\phi_{M}\equiv \left(\phi_{N}+\phi_{\lambda}\right)$ and $\phi_{N}\equiv-\frac{GM}{r}$ is the Newtonian potential. Substituting (\ref{B}) back into (\ref{newton limit3}), one obtains
\be
\frac{1}{2}\left(\frac{dr}{dt}\right)^2+\frac{J^2}{2 r^2}+ \phi_M=\frac{1}{2}\ ,
\label{potential}
\ee
where the effective Newtonian potential $\phi_{M}$ is given as
\be
\label{modified potential2}
\phi_M=-\frac{GM}{r}-\frac{GM}{r_s}\left(1+\frac{r}{r_e}\right)e^{-\frac{r_e}{r}}\, .
\ee

\vspace{8mm}
We are grateful to Y.-G. Jiang and S. Wang for useful discussions. This work was supported by the National Natural Science Fund of China under Grant No. 10875129 and No. 11075166.


\end{document}